\documentclass[aps,preprint,nofootinbib,superscriptaddress,prc]{revtex4}
\usepackage{amssymb}

\usepackage{epsfig}
\usepackage[bookmarksnumbered,bookmarksopen,colorlinks,citecolor=blue,linkcolor=blue]{hyperref}
\begin{document}


\title{Nuclear mass predictions with radial basis function approach}

\author{Ning Wang}
\email{wangning@gxnu.edu.cn} \affiliation{ Department of Physics,
Guangxi Normal University, Guilin 541004, People's Republic of
China }

\author{Min Liu}
 \affiliation{ Department of Physics,
Guangxi Normal University, Guilin 541004, People's Republic of
China }

\begin{abstract}
With the help of radial basis function (RBF) and the Garvey-Kelson
relation, the accuracy and predictive power of some global nuclear
mass models are significantly improved. The rms deviation between
predictions from four models and 2149 known masses falls to about
200 keV. The AME95-03 and AME03-Border tests show that the RBF
approach is a very useful tool for further improving the
reliability of mass models. Simultaneously, the differences from
different model predictions for unknown masses are remarkably
reduced and the isospin symmetry is better represented when the
RBF extrapolation is combined.
\end{abstract}

\maketitle

Nuclear mass predictions and evaluations are of great importance,
not only for various applications but also for test and
development of nuclear theory. The unmeasured masses are usually
predicted by using some global nuclear mass models in which some
physics are considered and the model parameters are determined by
the known masses \cite{Audi95,Audi} or by adopting some local mass
relations based on the measured masses of its neighbors. Some
global nuclear mass models such as the  finite range droplet model
(FRDM) \cite{Moll95}, the Weizs\"acker-Skyrme (WS) mass model
\cite{Wang,Wang10,Wang11}, the Hartree-Fock-Bogoliubov (HFB) model
\cite{HFB17} and the Duflo-Zuker (DZ) mass model \cite{DZ28}
successfully reproduce the measured masses with accuracy at the
level of 300 to 600 keV. However, the divergence for describing
the masses of the extremely neutron-rich nuclei from these
different global mass models indicates that more physics and more
information about nuclear force should be considered in the
models. The uncertainty in nuclear force and the limiting of
computational resources cause great difficulties for further
improving these available global nuclear mass models. On the other
hand, one could use the local mass relations such as the isobaric
multiplet mass equation (IMME) \cite{Ormand}, the Garvey-Kelson
(GK) relations \cite{GK} and the residual proton-neutron
interactions \cite{Zhao,Zhao1,Zhao2} to give predictions of
unmeasured masses. It is found that when these local mass
relations are used to predict the masses of nuclei in an iterative
fashion, the intrinsic error grows rapidly \cite{Mora09} due to:
1) the local mass relations are just approximately satisfied in
known masses and 2) the previously predicted masses are used on
each new iteration and a systematic error accumulates (see Table 1
in \cite{Mora09} and the expressions $\sigma_{\rm pred}$ in
\cite{Zhao2}). To improve the accuracy of nuclear mass
predictions, the systematics of nuclear mass surface is analyzed
in the image reconstruction techniques \cite{Mora,Mora1} based on
the Fourier transform (CLEAN algorithm) by combining some global
nuclear mass models. Compared with other local mass relations
mentioned previously, the image reconstruction techniques predict
the mass of a unmeasured nucleus by using many more known masses
rather than just the masses of its neighbors. Therefore, more
information from the experimental data could be involved for the
mass predictions. It is found that important improvements in the
predictions given by the different models were obtained with the
CLEAN reconstruction.

In this work, we attempt to propose a more efficient systematic
method based on the radial basis function approach
\cite{Hardy,Buh} together with the available nuclear mass models
for further improving the nuclear mass predictions. The mass
predictions for unmeasured nuclei can be treated as a problem of
mass surface extrapolation from the scattered experimental data.
The most prominent global interpolation and extrapolation scheme
is the radial basis function (RBF) approach that originates from
Hardy's multiquadric interpolation \cite{Hardy}. As a powerful
solution to the problem of scattered data fitting, the radial
basis function is widely applied in surface reconstruction. The
simplest form of RBF solution is written as $S({\bf x})=\sum
\limits_{i=1}^{m} w_i \phi(\| {\bf x}-{\bf x}_i \|)$. Where, ${\bf
x}_i$ denotes the points from measurement, $w_i$ is the weight of
center ${\bf x}_i$, $\phi$ is the basis function, $\|{\bf x}\|$ is
the Euclidean norm and $m$ is the number of the scattered data to
be fitted. Given $m$ samples (${\bf x}_i, f_i$), one wishes to
reconstruct the smooth function $S({\bf x})$ with $S({\bf
x}_i)=f_i$. The RBF weights $w_i$ are determined by the solution
of the linear system resulting from the interpolation condition.
Standard basis functions include:
\begin{itemize}
 \item{Spline: $\phi(r)=r$, or $\phi(r)=r^2\log(r)$},
 \item{Gaussian: $\phi(r)=\exp(-c \, r^2)$, with $c>0$},
\item{Multiquadric: $\phi(r)=\sqrt{r^2+c^2}$},
\item{Inverse
multiquadric: $\phi(r)=1/\sqrt{r^2+c^2}$}.
\end{itemize}

With the RBF approach, the difference $R(N,Z)=M_{\rm exp}-M_{\rm
th}$ between the calculated masses $M_{\rm th}$ with global
nuclear mass models and the experimental data $M_{\rm exp}$ could
be reconstructed. Once the reconstructed function $S(N,Z)$ is
obtained, the revised masses for unmeasured nuclei are given by
$M_{\rm th}^{\rm RBF}=M_{\rm th}+S$. In this work, we perform
three tests for each mass model, and for all tests we only
consider nuclei with neutron number $N\ge8$ and proton number
$Z\ge 8$. The first one is that we reconstruct the function
$S(N,Z)$ for a selected known nucleus based on other known masses
together with a certain global mass model. In other words, we take
the 2148 known masses of nuclei for training the RBF ($m=2148$)
and use the remaining one nucleus from the 2149 nuclei in the
atomic mass evaluation of 2003 (AME2003) \cite{Audi} as test. The
corresponding results from this kind of cross-validation will be
shown in Fig. 1 and Table I. The second one is that we take the
masses in AME1995 \cite{Audi95} for training the RBF ($m=1760$)
and predict the 389 "new" masses in AME2003, and the corresponding
results will be listed in Table II. The third one is that we take
the masses of nuclei near the $\beta$-stability line (nuclei with
neutron separation energy of $5\le S_n \le 12$ MeV) for training
the RBF ($m=1700$) and predict the remaining 449 masses of nuclei
approaching drip lines, and the results will be shown in Table
III. We find that the mass deviation $R(N,Z)$ can be reconstructed
relatively better with $\phi(r)=r$, i.e., a natural spline
function. Therefore, we adopt the basis function $\phi(r)=r$ in
the calculations.

\begin{figure}
\includegraphics[angle=-0,width= 0.7\textwidth]{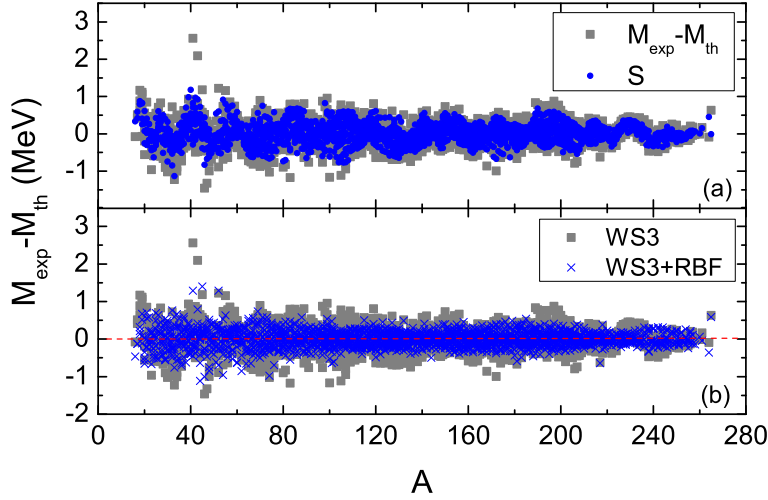}
 \caption{(Color online) Difference between predicted masses with
 the WS3 model and the experimental data. The solid circles in (a) denote the reconstructed function $S(N,Z)$ from the radial basis function with $\phi=r$.
 The crosses in (b) denote the corresponding deviations from data when the function $S$ is added to the masses with WS3. }
\end{figure}

\begin{table}
\caption{ rms $\sigma$ deviations between 2149 know masses
\cite{Audi} and predictions of five models (in keV). Here, to
predict the mass of a nucleus in AME2003 we take the remaining
2148 known masses for training the RBF ($m=2148$), see text for
details.}
\begin{tabular}{cccc}
 \hline\hline
       & ~~~ Model ~~~& Model+RBF ~~~& Model+RBF+GK12\\
\hline
 WS3 \cite{Wang11}       & 336 & 223 & 184\\
 DZ28 \cite{DZ28}        & 360 & 227 & 187\\
 FRDM \cite{Moll95}      & 656 & 283 & 216\\
 HFB17 \cite{HFB17}      & 581 & 390 & 313\\
 WS*  \cite{Wang10}      & 441 & 256 & 217\\
 \hline\hline
\end{tabular}
\end{table}

\begin{figure}
\includegraphics[angle=-0,width= 0.7\textwidth]{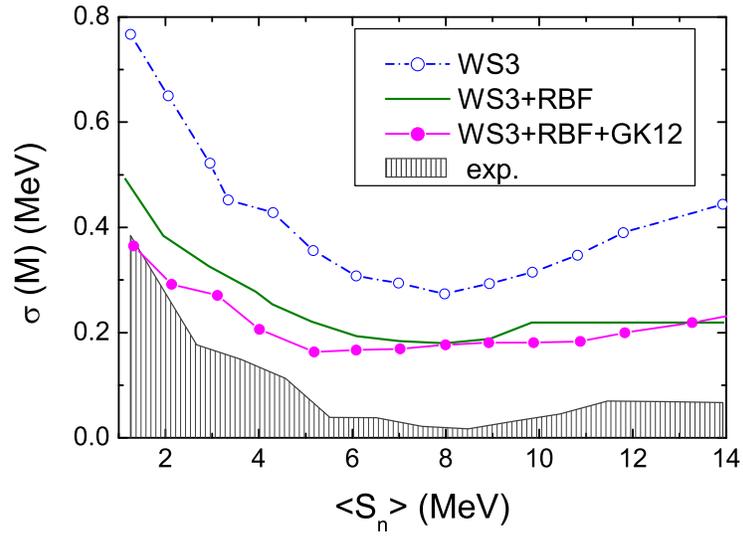}
 \caption{(Color online) rms deviation with respect to the masses as a function of average neutron-separation energy of nuclei.
 The shades present the average standard deviation errors of the measured masses
 \cite{Audi}.
}
\end{figure}
\begin{figure}
\includegraphics[angle=-0,width= 0.7\textwidth]{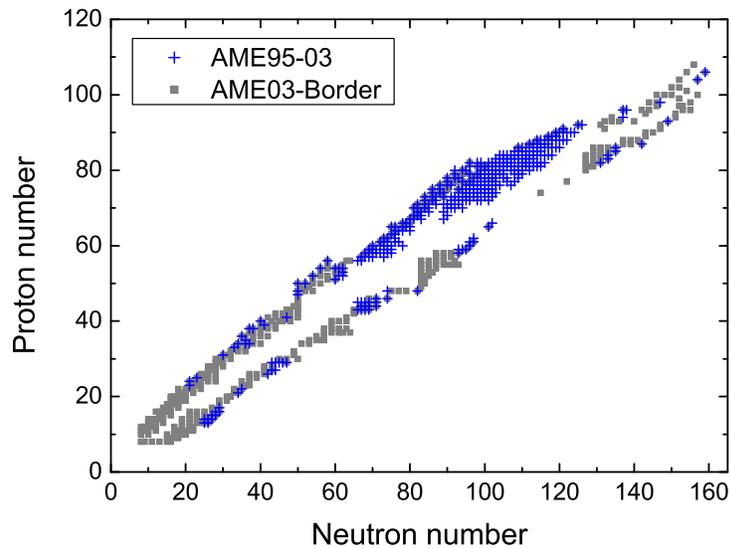}
 \caption{(Color online) Positions of nuclei predicted in the AME95-03 test (crosses) and the AME03-Border test (squares).
}
\end{figure}

In Fig. 1(a), we show the differences between the calculated
masses with WS3 model \cite{Wang11} and the experimental data
\cite{Audi} (gray squares). The reconstructed function $S$ (solid
circles) is also shown for comparison. One sees that the RBF
reproduces the differences $M_{\rm exp}-M_{\rm th}$ with high
quality and has good approximation properties. With the
approximating function $S(N,Z)$, the differences between the
predicted masses and the experimental data are dramatically
reduced [see the crosses in Fig. 1(b)]. In Table I, we list the
rms deviations between the 2149 know masses \cite{Audi} and
predictions from five mass models: WS3 \cite{Wang11}, DZ28
\cite{DZ28}, FRDM \cite{Moll95}, HFB17 \cite{HFB17} and WS*
\cite{Wang10}. The column Model+RBF means the reconstructed
function $S$ with the RBF approach is added to the calculated
masses with the five models. The column Model+RBF+GK12 denotes
that the Garvey-Kelson relation \cite{GK}, which contains 12
estimates for a nucleus with the corresponding values of its 21
neighbors, is also adopted for further improving the smoothness of
the function $S(N,Z)$. With the help of the RBF and the GK
relation, the rms deviations from the 2149 nuclei are reduced
sharply for all five models, and the results from four models
reach around 200 keV. In Fig. 2, we show the rms deviations
obtained with the first test for WS3 but as a function of average
neutron-separation energy of nuclei. With the RBF approach, the
rms deviations are reduced obviously, especially for nuclei
approaching the drip lines.

\begin{table}
\caption{ rms deviations with respect to 389 "new" masses
  in AME2003 based on the mass models and the measured
masses in AME1995 ($m=1760$) for training the RBF (in keV). }
\begin{tabular}{ccc}
 \hline\hline
       & ~~~ Model ~~~& Model+RBF \\
\hline
 WS3        & 378 & 311\\
 DZ28       & 430 & 341\\
 FRDM       & 536 & 351\\
 HFB17      & 519 & 380\\
 WS*        & 517 & 358\\
 \hline\hline
\end{tabular}
\end{table}

The second test, i.e., AME95-03 test, is usually used to check the
predictive power of mass models.  The crosses in Fig. 3 denote the
positions of nuclei to be predicted in the AME95-03 test. Table II
lists the rms deviations with respect to 389 "new" masses in
AME2003 based on the five mass models and the measured masses in
AME1995 for training. The reduction of rms deviation is $18\%$ for
the WS3 model, $21\%$ for the DZ28 model, $35\%$ for the FRDM,
$27\%$ for the HFB17 model and $31\%$ for the WS* model,
respectively, when the RBF approach is combined. We note that the
reduction of the rms deviation ($N,Z\ge 8$) is about $12\%$ with
the CLEAN reconstruction \cite{Mora1} combining the 31-parameter
Duflo-Zuker (DZ31) mass model \cite{DZ31} in this test. The
corresponding result with the RBF approach remarkably reaches
about $23\%$. Combining the liquid drop model (LDM) mentioned in
\cite{Mora}, the rms reduction in the AME95-03 test reaches
$\sim$$54\%$ with the CLEAN reconstruction and $\sim$$72\%$ with
the RBF approach, respectively. It seems that the radial basis
function approach is a more efficient tool for improving the
accuracy of nuclear mass predictions. Furthermore, in the CLEAN
algorithm one needs to perform a series of iterations until a
given stopping criteria ($\sigma$ =100 keV for example) which is
not required in the RBF approach. In the RBF approach, one just
needs to calculate the weights $w_i$ which can be estimated using
the matrix methods of linear least squares, because the
approximating function $S(N,Z)$ is linear in the weights.

\begin{table}
\caption{ rms deviations (in keV) from 449 masses of nuclei
  approaching AME03-Border based on the mass models and
the known masses of nuclei with neutron separation energy of $5\le
S_n \le 12$ MeV ($m=1700$) for training the RBF. }
\begin{tabular}{ccc}
 \hline\hline
       & ~~~ Model ~~~& Model+RBF \\
\hline
 WS3        & 423  & 367\\
 DZ28       & 491  & 392\\
 FRDM       & 855  & 582\\
 HFB17      & 730  & 575\\
 WS*        & 591  & 417\\
 \hline\hline
\end{tabular}
\end{table}

In Table III, we list the rms deviations with respect to the 449
masses of nuclei approaching AME03-Border based on the five models
and the known masses of nuclei with neutron separation energy of
$5\le S_n \le 12$ MeV in AME2003 for training. The definition of
the AME03-Border test here is slightly different from that in
\cite{Mora1} (see the squares in Fig. 3).  The rms deviation is
reduced by $20\%$ for the DZ28 model, $21\%$ for the HFB17 model,
$29\%$ for the WS* model and $32\%$ for the FRDM, respectively.
For the WS3 model, we obtain the reduction of $13\%$. These
calculations indicate that the reliability of the available mass
models can be significantly improved by combining the RBF
approach. With the GK relation, the results in Table II and III
can be further improved as those do in Table I, because the GK
relation is also well satisfied at the mass region with nuclei far
from the $\beta$-stability line \cite{GK}.

\begin{figure}
\includegraphics[angle=-0,width= 0.7\textwidth]{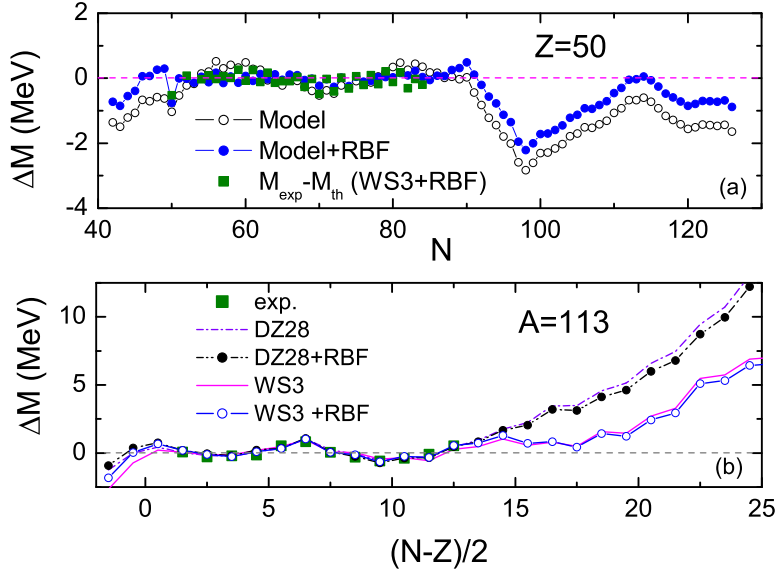}
 \caption{(Color online) (a) Difference between the calculated masses with the mass model
 WS3 and those with DZ28 for Sn isotopes. The solid circles denote
 the corresponding results when the RBF approach is combined. (b) Difference of the calculated masses with the mass
 models from the results through fitting the experimental masses
 with a parabola for isobaric nuclei of $A=113$. The squares
 denote the deviations from the experimental data, i.e. $M_{\rm exp}-M_{\rm
 fit}$. The solid and the dash-dotted curve denote the results
 with
 WS3 and DZ28, respectively.
}
\end{figure}

In addition, we study the predictions from these mass models for
unmeasured nuclei. It is known that the predictions from different
models towards the neutron-drip line tend to diverge. As an
example, the difference $\Delta M=M ({\rm DZ28}) -M ({\rm WS3})$
between the calculated masses with the DZ28 model and those with
the WS3 model for Sn isotopes is shown in Fig. 4(a) (open circles)
as a function of neutron numbers. For known nuclei, the
differences between the calculated masses from the two models are
small. However, the deviations reach a few MeV for nuclei
approaching the drip lines. The solid circles in Fig. 4(a) denote
the corresponding results when the RBF approach is combined. The
differences between the predicted masses from the two models are
reduced by about one MeV. Simultaneously, we list in Table IV the
rms deviations with respect to the evaluated masses in AME2003
\cite{Audi} (marked by $\#$) and those in \cite{Zhao2} based on
the residual proton-neutron interactions. We find that when the
RBF approach ($m=2149$) and the GK relation are involved in the
mass predictions, the rms deviations with respect to the evaluated
masses in AME2003 are remarkably reduced. The rms reductions reach
$23\%$ for the WS3 model, $34\%$ for the DZ28 model, $38\%$ for
the FRDM, $23\%$ for the HFB17 model and $31\%$ for the WS* model,
respectively. The rms deviations from the evaluated masses based
on the residual proton-neutron interactions \cite{Zhao2} are also
significantly reduced, with rms reduction of $38\%$ for the WS3
model, $42\%$ for the DZ28 model, $44\%$ for the FRDM, $33\%$ for
the HFB17 model and $46\%$ for the WS* model, respectively. These
calculations also demonstrate that the differences from these
different models can be remarkably reduced when the RBF approach
and the GK relation are applied.

\begin{table}
\caption{rms deviations (in MeV) with respect to the evaluated
 masses in AME2003 \cite{Audi} (marked by $\#$) and those in
 \cite{Zhao2} for nuclei with $Z\le102$
based on the residual proton-neutron interactions. Here the RBF
approach ($m=2149$) and the GK relation are involved in the mass
predictions with the five models. $n_1$ and $n_2$ denote the
number of evaluated masses taken from AME2003 and \cite{Zhao2} for
the rms calculations, respectively. }
\begin{tabular}{ccccc}
 \hline\hline
       & ~~~$n_1$~~~ &   AME2003 \cite{Audi} & ~~~$n_2$~~~ & ~~~$\delta V_{1p1n}$ \cite{Zhao2} ~~~ \\
\hline
 WS3   & 935 &0.629   & 459 & 0.479\\
 DZ28  & 935 &1.058   & 459 & 0.470\\
 FRDM  & 935 &0.784   & 459 & 0.597\\
 HFB17 & 892 &0.743   & 459 & 0.706\\
 WS*   & 935 &0.626   & 459 & 0.498\\
 \hline\hline
\end{tabular}
 \end{table}

In Fig. 4(b) we show the difference $\Delta M=M_{\rm th}-M_{\rm
fit}$ between the predicted masses with models and the results by
fitting the known masses with a parabola for a series of isobaric
nuclei with $A=113$. According to the isobaric multiplet mass
equation (IMME) which is a basic prediction leading from the
isospin concept, the masses of isobaric nuclei can be expressed as
$M_{\rm fit}(A,T_z)=a+b T_z+c T_z^2$ with $T_z= (N-Z)/2$. One sees
from Fig. 4(b) that the deviation $M_{\rm th}-M_{\rm fit}$ is
quite large for nuclei approaching the neutron drip line. When the
RBF approach is combined, the deviations are slightly reduced,
which indicates that the isospin symmetry is relatively better
represented with the help of the RBF extrapolation.
Simultaneously, we study the $b$ coefficients in the IMME with
different models. The $b$ coefficients in the IMME for nuclei can
be extracted from the binding energies of pairs of mirror nuclei
$b=\frac{BE(T=T_z)-BE(T=-T_z)}{2T}$ with the isospin $T=|N-Z|/2$.
We calculate rms deviations between the experimental $b$
coefficients in the IMME for 62 pairs of mirror nuclei and the
predicted ones with the models. When the RBF approach and the GK
relation are used, the rms deviations from the experimental $b$
coefficients are significantly reduced from 159 keV to 97 keV for
the WS3 model, from 274 keV to 144 keV for the FRDM and from 180
keV to 91 keV for the WS* model, respectively. It implies that the
isospin symmetry is an important concept for constraining the
nuclear mass models.

In summary, we proposed an efficient systematic method based on
the radial basis function (RBF) approach for improving the
accuracy and predictive power of global nuclear mass models. With
the help of RBF and the Garvey-Kelson relation, the rms deviation
between the predictions from four mass models and the 2149 known
masses falls to about 200 keV. The AME95-03 and  AME03-Border
tests show that the RBF approach provides us with a very useful
tool, even more efficient than the CLEAN algorithm, for further
reducing the rms deviation from the experimental data. In
addition, with the RBF extrapolation, the divergence from
different model predictions for unknown masses is remarkably
improved, and the isospin symmetry is better represented.

\begin{center}
\textbf{ACKNOWLEDGEMENTS}
\end{center}

This work was supported by National Natural Science Foundation of
China, Nos 10875031, 10847004, 11005022 and 10979024. The obtained
mass tables with the radial basis function approach are available
on line \cite{RBF}.

\end{document}